\documentclass[nofootinbib,preprintnumbers,prd,superscriptaddress,twocolumn,showpacs]{revtex4-1}

\usepackage{amsmath}
\usepackage{amsfonts}
\usepackage{amssymb}
\usepackage{graphicx}
\usepackage[titletoc]{appendix}
\usepackage{color}
\usepackage{hyperref}
\usepackage{cleveref}
\usepackage[rightcaption]{sidecap}
\usepackage{subfigure}
\usepackage{comment}
\usepackage{soul}
\usepackage{cancel}
\usepackage{dcolumn}

\usepackage{array}
\usepackage{ctable}
\usepackage{multirow}
\usepackage{siunitx}
\usepackage{longtable}
\usepackage{tabularx}
\usepackage{booktabs}
\usepackage{float}

\graphicspath{{Graphics/}}

\def\be{\begin{equation}}
\def\ee{\end{equation}}
\def\bea{\begin{eqnarray}}
\def\eea{\end{eqnarray}}

\def\prd{Phys. Rev. D}
\def\mnras{MNRAS}
\def\aj{AJ}
\def\apj{ApJ}
\def\apjl{ApJ Lett.}
\def\apjs{ApJ Suppl. Ser.}

\def\aap{A\&A}

\def\araa{Annual Rev. of Astron. Astrophys.}

\def\jcap{JCAP}

\definecolor{vividviolet}{rgb}{0.62, 0.0, 1.0}
\definecolor{amaranth}{rgb}{0.9, 0.17, 0.31}
\definecolor{palatinateblue}{rgb}{0.15, 0.23, 0.89}
\definecolor{brightpink}{rgb}{1.0, 0.0, 0.5}
\definecolor{cornflowerblue}{rgb}{0.39, 0.58, 0.93}
\definecolor{deepcarminepink}{rgb}{0.94, 0.19, 0.22}
\definecolor{radicalred}{rgb}{1.0, 0.21, 0.37}

\hypersetup{ linktoc=all,
    colorlinks, linkcolor={palatinateblue},
    citecolor={brightpink}, urlcolor={amaranth}
}

\DeclareUnicodeCharacter{2212}{\ensuremath{-}}

\begin{document}

\title{Breaking matter degeneracy in a model-independent way through the Sunyaev-Zeldovich effect}

\author{Anna Chiara Alfano}
\email{annac.alfano@studenti.unina.it}
\affiliation{Scuola Superiore Meridionale, Largo S. Marcellino 10, 80138 Napoli, Italy.}

\author{Orlando Luongo}
\email{orlando.luongo@unicam.it}
\affiliation{Universit\`a di Camerino, Divisione di Fisica, Via Madonna delle carceri 9, 62032 Camerino, Italy.}
\affiliation{SUNY Polytechnic Institute, 13502 Utica, New York, USA.}
\affiliation{INFN, Sezione di Perugia, Perugia, 06123, Italy.}
\affiliation{INAF - Osservatorio Astronomico di Brera, Milano, Italy.}
\affiliation{Al-Farabi Kazakh National University, Al-Farabi av. 71, 050040 Almaty, Kazakhstan.}

\author{Marco Muccino}
\email{marco.muccino@lnf.infn.it}
\affiliation{Universit\`a di Camerino, Divisione di Fisica, Via Madonna delle carceri 9, 62032 Camerino, Italy.}
\affiliation{Al-Farabi Kazakh National University, Al-Farabi av. 71, 050040 Almaty, Kazakhstan.}

\date{\today}

\begin{abstract}
We propose a model-independent \textit{Bézier parametric interpolation} to alleviate the degeneracy between baryonic and dark matter abundances by means of intermediate-redshift data. To do so, we first interpolate the observational Hubble data to extract cosmic bounds over the (reduced) Hubble constant, $h_0$, and interpolate the angular diameter distances, $D(z)$, of the galaxy clusters, inferred from the Sunyaev-Zeldovich effect, constraining the spatial curvature, $\Omega_k$. Through the so-determined Hubble points and $D(z)$, we interpolate uncorrelated data of baryonic acoustic oscillations bounding the baryon ($\omega_b = h^2_0\Omega_b$) and total matter ($\omega_m = h^2_0\Omega_m$) densities, reinforcing the constraints on $h_0$ and $\Omega_k$ with the same technique. Instead of pursuing the usual treatment to fix $\omega_b$ via the value obtained from the cosmic microwave background to remove the matter sector degeneracy, we here interpolate the acoustic parameter from correlated baryonic acoustic oscillations. The results of our Monte Carlo--Markov chain simulations turn out to agree at $1$--$\sigma$ confidence level with the flat $\Lambda$CDM model. While our findings are roughly suitable at $1$--$\sigma$ with its non-flat extension too, the Hubble constant appears in tension up to the $2$--$\sigma$ confidence level. Accordingly, we also reanalyze the Hubble tension with our treatment and find  our expectations slightly match local constraints.
\end{abstract}



\maketitle

\section{Introduction}

In the standard cosmological puzzle, the accelerated expansion of the universe represents a solid evidence, first detected by means of type Ia supernovae (SNe Ia) observations \cite{1998AJ....116.1009R,1999ApJ...517..565P}. The presence of dust under the form of baryons and (cold) dark matter alone is not capable of accelerating the universe today\footnote{Remarkable exceptions are unified models of dark energy, in which dark energy emerges as a consequence of dark matter, see e.g. Ref.~\cite{Boshkayev:2019qcx} and references therein.}. Accordingly, the acceleration of the universe can be attributed to an exotic \emph{dark energy} fluid, exerting a repulsive gravitational effect due to its negative equation of state \cite{2000IJMPD...9..373S,2006IJMPD..15.1753C,tsujikawa2011dark}.

The universe energy momentum budget yields a dark energy contributing for $\sim 70\%$, while the remaining $\sim 30 \%$ consists of cold dark matter and baryonic matter \cite{Planck2018}. Within the current cosmological background scenario, the $\Lambda$CDM model, it is not possible to measure separately baryonic matter and cold dark matter by relying solely on SNe Ia observations. This limitation is not only a characteristic of the current cosmological background scenario, where dark energy is under the form of a cosmological constant $\Lambda$ \cite{2001LRR.....4....1C}, but of any dark energy scenario. However, among all possible frameworks, the $\Lambda$CDM model remains the most statistically favored approach to describe the universe large scale dynamics, albeit plagued by conceptual inconsistencies.

Specifically, the $\Lambda$CDM paradigm still suffers from the coincidence and fine-tuning problems \cite{2006IJMPD..15.1753C} and cosmological tensions related to the discrepancies in Hubble constant measurements and to the amplitude of the clustering of matter $S_8$ using high and low-redshift probes \cite{bernalcosmic, 2021CQGra..38o3001D, 2022JHEAp..34...49A}. To overcome these issues, several theoretical efforts have been made to find alternative models that reproduce the successful features of the $\Lambda$CDM paradigm \cite{nostro,2022CQGra..39s5014D,mio2022}.

From a phenomenological standpoint, diverse model-independent approaches have been proposed in order to estimate the cosmological parameters without assuming \textit{a priori} the  cosmological model \cite{2013Galax...1..216C, 2016IJGMM..1330002D, 2021MNRAS.503.4581L, 2023MNRAS.518.2247L, 2007MNRAS.380.1573S, 2010PhRvD..81h3537S, 2018JCAP...10..015H}.

Motivated by the need of \emph{disentangling} baryons from cold dark matter, we here propose a model-independent approach based on the \emph{B\'ezier parametric interpolation} that: a) approximates the Hubble rate $H(z)$ via the observational Hubble data (OHD) \cite{2019MNRAS.486L..46A} and constrains the Hubble constant, and b) uses this approximation to interpolate the angular diameter distances $D(z)$ of galaxy clusters, inferred from the Sunyaev-Zeldovich (SZ) effect, to provide bounds on the spatial curvature $\Omega_k$.
Thus, by virtue of the above extrapolations, we fit both uncorrelated and correlated data sets of baryonic acoustic oscillations (BAO) to constrain the matter densities for baryons $\Omega_b$ and for all matter components $\Omega_m$, via $\omega_b=h_0^2\Omega_b$ and $\omega_m=h_0^2\Omega_m$,  which the comoving sound horizon, $r_s$, depends upon \cite{2021PhRvD.104d3521A}.
We point out that the correlated BAO turn out to be  essential to break the degeneracy between $\omega_m$ and $\omega_b$ \cite{1999MNRAS.304...75E}, differently of the standard procedure fixing $\omega_b$ to  cosmic microwave background (CMB) value  \cite{Planck2018}. Accordingly, we intentionally put aside the CMB data from \textit{Planck} satellite and the \textit{Pantheon} catalog of SNe Ia \cite{2018ApJ...859..101S}, due to their tension in determining $h_0$. The results of our Monte Carlo--Markov chain (MCMC) simulations, based on the Metropolis-Hastings algorithm, suggest that the  matter sector degeneracy is likely alleviated without the need for CMB and/or SNe Ia data. Notably, this technique, while valuable, does not completely eliminate the cosmic tensions that still persist. In fact, in the flat scenario, our $h_0$ turns out to be in good agreement with the Planck estimates \cite{Planck2018}, but still barely consistent with SNe Ia \cite{2022ApJ...934L...7R}, at $1$--$\sigma$ confidence level, while in the non-flat case, our $h_0$ is consistent with Planck only at $2$--$\sigma$ confidence level \cite{Planck2018}, indicating that spatial curvature may influence the $h_0$ measurements. Consequently, when comparing our results with expectations based on the flat (non-flat) $\Lambda$CDM framework, we obtain constraints that are in good agreement with the standard cosmological model at $1$--$\sigma$ ($2$--$\sigma$) level, certifying that the $\Lambda$CDM paradigm is still supported within our treatment.

The paper is structured as follows. In Sect.~\ref{sec2}, we introduce the data sets and show how to use them to perform model-independent interpolations through B\'ezier curves. In Sect.~\ref{sec3}, we show the numerical constraints on $h_0$, $\Omega_k$, $\omega_b$ and $\omega_m$, obtained from our MCMC analysis, and compare them with the best-fit parameters inferred from both the flat and the non-flat $\Lambda$CDM models. In Sect.~\ref{sec4}, we summarize the physical implications of our efforts, reporting our conclusions and perspectives.

\section{Methods}\label{sec2}

To obtain model-independent estimates on the key cosmological parameters, we utilize intermediate redshift catalogs, such as OHD, SZ and BAO data sets.
The \emph{Pantheon} catalog of SNe Ia \cite{2018ApJ...859..101S} and CMB data from the \emph{Planck} satellite \cite{Planck2018} are intentionally excluded in view of the existing tension on their estimates on $h_0$ \cite{2022ApJ...934L...7R,Planck2018}.

The model-independent technique here resorted is based on the well-established B\'ezier parametric interpolation, firstly introduced in Ref.~\cite{LM2020} and widely adopted in cosmological contexts with promising results \cite{2019MNRAS.486L..46A,2021MNRAS.501.3515M,2023MNRAS.518.2247L,2023MNRAS.523.4938M}.
We employ this technique to
\begin{itemize}
    \item[-] interpolate the Hubble rate $H(z)$ from OHD \cite{2002ApJ...573...37J},
    \item[-] use it to derive the angular diameter distance $D(z)$, on which SZ and BAO observables are based, and
    \item[-] interpolate BAO data to break the baryonic--dark matter degeneracy.
\end{itemize}
It is worth stressing that the so-interpolated $H(z)$ and $D(z)$ bear no \emph{a priori} assumptions on $\Omega_k$, as long as the data sets do not carry specific priors on it.

\subsection{Hubble rate model-independent reconstruction via B\'ezier polynomials}

We interpolate the most updated sample of $N_{\rm O}=33$ Hubble rate measurements, spanning up to the maximum redshift $z_{\rm m}=1.965$ (see Tab.~\ref{tab:OHD}).
These measurements are determined from spectroscopic measurements of the differences in age $\Delta t$ and redshift $\Delta z$ of couples of passively evolving galaxies (formed at the same time) by resorting
the identity $H(z)=-(1+z)^{-1}\Delta z/\Delta t$ \cite{2002ApJ...573...37J}.
\begin{table}
\centering
\setlength{\tabcolsep}{2.4em}
\renewcommand{\arraystretch}{1.1}
   \begin{tabular}{lcc}
   \hline\hline
    $z$     &$H(z)$ &  Refs. \\
            &(km\,s$^{-1}$Mpc$^{-1}$)&\\
    \hline
    0.0708  & $69.0  \pm 19.6$ & \cite{Zhang2014} \\
    0.09    & $69.0  \pm 12.0$  & \cite{Jimenez2002} \\
    0.12    & $68.6  \pm 26.2$  & \cite{Zhang2014} \\
    0.17    & $83.0  \pm 8.0$   & \cite{Simon2005} \\
    0.179   & $75.0  \pm 4.0$   & \cite{Moresco2012} \\
    0.199   & $75.0  \pm 5.0$   & \cite{Moresco2012} \\
    0.20    & $72.9  \pm 29.6$  & \cite{Zhang2014} \\
    0.27    & $77.0  \pm 14.0$  & \cite{Simon2005} \\
    0.28    & $88.8  \pm 36.6$  & \cite{Zhang2014} \\
    0.352   & $83.0  \pm 14.0$  & \cite{Moresco2016} \\
    0.3802  & $83.0  \pm 13.5$  & \cite{Moresco2016} \\
    0.4     & $95.0  \pm 17.0$  & \cite{Simon2005} \\
    0.4004  & $77.0  \pm 10.2$  & \cite{Moresco2016} \\
    0.4247  & $87.1  \pm 11.2$  & \cite{Moresco2016} \\
    0.4497  & $92.8  \pm 12.9$  & \cite{Moresco2016} \\
    0.47    & $89.0\pm23.0$     & \cite{2017MNRAS.467.3239R}\\
    0.4783  & $80.9  \pm 9.0$   & \cite{Moresco2016} \\
    0.48    & $97.0  \pm 62.0$  & \cite{Stern2010} \\
    0.593   & $104.0 \pm 13.0$  & \cite{Moresco2012} \\
    0.68    & $92.0  \pm 8.0$   & \cite{Moresco2012} \\
    0.75    & $98.8\pm33.6$     & \cite{2022ApJ...928L...4B}\\
    0.781   & $105.0 \pm 12.0$  & \cite{Moresco2012} \\
    0.80    & $113.1\pm15.1$    & \cite{2023ApJS..265...48J}\\
    0.875   & $125.0 \pm 17.0$  & \cite{Moresco2012} \\
    0.88    & $90.0  \pm 40.0$  & \cite{Stern2010} \\
    0.9     & $117.0 \pm 23.0$  & \cite{Simon2005} \\
    1.037   & $154.0 \pm 20.0$  & \cite{Moresco2012} \\
    1.3     & $168.0 \pm 17.0$  & \cite{Simon2005} \\
    1.363   & $160.0 \pm 33.6$  & \cite{Moresco2015} \\
    1.43    & $177.0 \pm 18.0$  & \cite{Simon2005} \\
    1.53    & $140.0 \pm 14.0$  & \cite{Simon2005} \\
    1.75    & $202.0 \pm 40.0$  & \cite{Simon2005} \\
    1.965   & $186.5 \pm 50.4$  & \cite{Moresco2015} \\
\hline
\end{tabular}
\caption{The redshift distribution (first column) of the OHD measurements with the statistical errors (second column) and the reference papers (third column).}
\label{tab:OHD}
\end{table}

The best-fit, non-linear and monotonic growing function with the redshift of these data is a second order B\'ezier curve with coefficients $\alpha_i$ \cite{2019MNRAS.486L..46A,LM2020,2021MNRAS.501.3515M,2023MNRAS.518.2247L,2023MNRAS.523.4938M}, i.e.,
\begin{equation}
\label{bezier1}
H_2(z) = g_\alpha \sum_{i=0}^{2} \frac{2\alpha_i}{i!(2-i)!} \left(\frac{z}{z_{\rm m}}\right)^i \left(1-\frac{z}{z_{\rm m}}\right)^{2-i}\,,
\end{equation}
where we defined $g_\alpha=100\,{\rm km\,s}^{-1}{\rm Mpc}^{-1}$. This interpolated function can be also extrapolated beyond $z_{\rm m}$, to cover the redshift range of the other intermediate redshift probes that will be introduced in the following.

Fitting the OHD measurements with Eq.~\eqref{bezier1} provides a model-independent estimate of the dimensionless Hubble constant, since at $z=0$ it holds $H_0/g_\alpha\equiv h_0\equiv \alpha_0$.

For Gaussian distributed errors $\sigma_{H_k}$, the coefficients $\alpha_i$ are found by maximizing the log-likelihood function
\begin{equation}
\label{loglikeOHD}
    \ln \mathcal{L}_{\rm O} = -\frac{1}{2} \sum_{j=1}^{N_{\rm O}}\left\{\left[\dfrac{H_j-H_2(z_j)}{\sigma_{H_j}}\right]^2 + \ln(2\pi\sigma^2_{H_j})\right\}\,.
\end{equation}

\subsection{Constraining the curvature with SZ data}

The SZ effect is the spectral distortion of the CMB photons via inverse Compton scattering by high-energy electrons gas in galaxy clusters \cite{2002ARA&A..40..643C}.
While CMB photons is observed at microwave frequencies, intra-cluster electrons and their distribution are observed in the X-rays.

Combining the above observations enables the determination, for relatively high threshold signal-to-noise ratio, of the triaxial structure of the clusters and, thus, their morphology-corrected diameter angular distances $D(z)$. A sample of $N_{\rm S}=25$ clusters with such determined distances \cite{2005ApJ...625..108D} is listed in Tab.~\ref{tab:SZ}.

\begin{table}
\centering
\setlength{\tabcolsep}{3.em}
\renewcommand{\arraystretch}{1.1}
\begin{tabular}{lc}
\hline\hline
$z$     & $D(z)$ \\
        & (Mpc)\\
\hline
$0.023$	& $103\pm42$\\
$0.058$	& $242\pm61$\\
$0.072$ & $165\pm45$\\
$0.074$ & $369\pm62$\\
$0.084$	& $749\pm385$\\
$0.088$	& $448\pm185$\\
$0.091$	& $335\pm70$\\
$0.142$	& $478\pm126$\\
$0.176$	& $809\pm263$\\
$0.182$	& $451\pm189$\\
$0.183$ & $604\pm84$\\
$0.202$	& $387\pm141$\\
$0.202$	& $806\pm163$\\
$0.217$	& $1465\pm407$\\
$0.224$	& $1118\pm283$\\
$0.252$	& $946\pm131$\\
$0.282$	& $1099\pm308$\\
$0.288$	& $934\pm331$\\
$0.322$	& $885\pm207$\\
$0.327$	& $697\pm183$\\
$0.375$	& $1231\pm441$\\
$0.451$	& $1166\pm262$\\
$0.541$	& $1635\pm391$\\
$0.550$	& $1073\pm238$\\
$0.784$	& $2479\pm1023$\\
\hline
\end{tabular}
\caption{The sample of galaxy clusters with the redshift (first column) and the diameter angular distances (second column) as inferred from the SZ effect \cite{2005ApJ...625..108D}.}
\label{tab:SZ}
\end{table}

Using Eq.~\eqref{bezier1}, we obtain an interpolated angular diameter distance defined as
\begin{equation}
\label{eq:da2}
D_2(z) = \frac{c\left(1+z\right)^{-1}}{g_\alpha\alpha_0\sqrt{\Omega_k}} \sinh \left[\int_0^z \frac{g_\alpha\alpha_0\sqrt{\Omega_k} dz^\prime}{H_2(z^\prime)}\right]\ ,
\end{equation}
where $\sinh(\sqrt{\Omega_k}x)/\sqrt{\Omega_k}$ holds for a curvature parameter $\Omega_k>0$, becomes $\sin(\sqrt{|\Omega_k|}x)/\sqrt{|\Omega_k|}$ for $\Omega_k<0$, and reduces to $x$ for $\Omega_k=0$.

Again, for Gaussian distributed errors $\sigma_{D_j}$, $\alpha_i$ and $\Omega_k$ can be derived by maximizing the log-likelihood function
\begin{equation}
\label{loglikeSZ}
    \ln \mathcal{L}_{\rm S} = -\frac{1}{2} \sum_{j=1}^{N_{\rm S}}\left\{\left[\dfrac{D_j-D_2(z_j)}{\sigma_{D_j}}\right]^2 + \ln(2\pi\sigma^2_{D_j})\right\}\,.
\end{equation}

\subsection{Constraining the matter density with BAO}

In the early universe, BAO are acoustic waves generated by the gravitational interaction between the photon-baryon fluid and inhomogeneities \cite{2008cosm.book.....W}.
During the {\it drag epoch}, baryons decoupled from photons and ``froze in'' at a scale equal to the sound horizon at the drag epoch redshift $z_\mathrm{d}$, i.e., $r_\mathrm{s}\equiv r(z_\mathrm{d})$.
Such a characteristic scale is a standard ruler embedded in the galaxy distribution \cite{2019JCAP...10..044C}.

A numerically-reconstructed expression for $r_\mathrm{s}$, that also includes massive neutrinos, is given by \cite{2021PhRvD.104d3521A}
\begin{equation}
\label{eq:neutrino}
r_\mathrm{s} = \frac{a_1~e^{a_2\left(a_3+\omega_{\nu}\right)^2}}{a_4~\omega_b^{a_5}+a_6~\omega_m^{a_7}+a_8\left(\omega_b\hspace{1mm}\omega_m\right)^{a_9}}~ \mathrm{Mpc}\,,
\end{equation}
and leads to an improvement in the accuracy, over the most used expression \cite{2015PhRvD..92l3516A}, of a factor $\sim3$ in the range within $3\sigma$ of the $\Lambda$CDM best-fit parameters \cite{Planck2018}, and a factor $\sim 30$ in the broader range \cite{2021PhRvD.104d3521A}.
In Eq.~\eqref{eq:neutrino}, the cosmological parameters for baryons $\omega_b$ and for both baryonic and dark matter $\omega_m$ are free parameters; for massive neutrino species we fixed $\omega_{\nu}=0.000645$ \cite{2015PhRvD..92l3516A}.
The other coefficients have the following values \cite{2021PhRvD.104d3521A}
\begin{equation}
\nonumber
\begin{array}{lll}
a_1 = 0.0034917, & a_2=-19.972694, & a_3=0.000336186\,,\\
a_4 = 0.0000305, & a_5=0.22752,    & a_6=0.0000314257\,,\\
a_7 = 0.5453798, & a_8=374.14994,  & a_9=4.022356899\,.\\
\end{array}
\end{equation}

To determine $r_\mathrm{s}$ from Eq.~\eqref{eq:neutrino} requires to break the degeneracy between $\omega_b$ and $\omega_m$ \cite{1999MNRAS.304...75E}.
In general, this is achieved by fixing $\omega_b$ with the value got from the CMB.
However, as stated above, CMB data are intentionally excluded because of the $H_0$ tension. Therefore, in the following we work out a procedure, based on different data sets of BAO measurements, that aims not only at providing constraints on $\omega_b$ and $\omega_m$ (and consequently on $r_\mathrm{s}$), but also at reinforcing the constraints on $h_0$ and $\Omega_k$ got from OHD and SZ data sets, respectively.

\begin{table}
\centering
\setlength{\tabcolsep}{.5em}
\renewcommand{\arraystretch}{1.1}
\begin{tabular}{llcc}
\hline\hline
Survey          & $z$       & $\delta(z)$
                &  Refs. \\
\hline
6dFGS		    & $0.106$	& $0.336\pm0.015$
                & \cite{2011MNRAS.416.3017B}\\
SDSS-DR7 	    & $0.15$    & $0.2239\pm0.0084$
                & \cite{2015MNRAS.449..835R}\\
SDSS-DR7	    & $0.2$ 	& $0.1905\pm0.0061$
                & \cite{2010MNRAS.401.2148P}\\
SDSS-DR11 LOWZ  & $0.32$	& $0.1181\pm0.0023$
                & \cite{2014MNRAS.441...24A}\\
SDSS-DR7 LRG	& $0.35$	& $0.1126\pm0.0022$
                & \cite{2012MNRAS.427.2132P}\\
BOSS DR12		& $0.38$    & $0.1001\pm0.0011$
                & \cite{2017MNRAS.470.2617A}\\
BOSS DR12		& $0.51$	& $0.0787\pm0.0008$
                & \cite{2017MNRAS.470.2617A}\\
SDSS-III/DR9 	& $0.57$    & $0.0726\pm0.0007$
                & \cite{2014MNRAS.441...24A}\\
BOSS DR12		& $0.61$    & $0.0691\pm0.0007$
                & \cite{2017MNRAS.470.2617A}\\
SDSS-IV/DR14	& $0.72$	& $0.0622\pm0.0016$
                & \cite{2018ApJ...863..110B}\\
SDSS-IV/DR14	& $1.52$	& $0.0385\pm0.0015$
                & \cite{2018MNRAS.473.4773A}\\
\hline\hline
Survey          & $z$       & $\Delta(z)$
                &  Refs. \\
\hline
SDSS-III DR8	& $0.54$	& $9.212\pm0.410$
                & \cite{2012ApJ...761...13S}\\
DECals DR8		& $0.697$	& $10.387\pm0.496$
                & \cite{2020ApJ...904...69S}\\
DES Year1		& $0.81$	& $10.75\pm0.43$
                & \cite{2019MNRAS.483.4866A}\\
DECals DR8		& $0.874$	& $11.372\pm0.693$
                & \cite{2020ApJ...904...69S}\\
\hline\hline
Survey          & $z$       & $\Theta(z)$
                &  Refs. \\
\hline
eBOSS DR16 BAO+RSD
                & $1.48$ 	& $13.23\pm0.47$
                & \cite{2021MNRAS.500.1201H}\\
eBOSS DR16		& $2.334$	& $8.99\pm0.19$
                & \cite{2020ApJ...901..153D}\\
                \hline\hline
Survey          & $z$       & $A(z)$
                &  Refs. \\
\hline
WiggleZ			& $0.44$  	& $0.474\pm0.034$
                & \cite{2012MNRAS.425..405B}\\
WiggleZ			& $0.6$ 	& $0.442\pm0.020$
                & \cite{2012MNRAS.425..405B}\\
WiggleZ			& $0.73$ 	& $0.424\pm0.021$
                & \cite{2012MNRAS.425..405B}\\
\hline
\end{tabular}
\caption{The four BAO catalogs with surveys (first column), redshifts (second column), measurements with errors (third column), and references (fourth column).}
\label{tab:BAO}
\end{table}

Tab.~\ref{tab:BAO} lists four kind of BAO data sets.
Resorting the interpolations in Eqs.~\eqref{bezier1} and \eqref{eq:da2}, the first set of $N_\delta=11$ measurements is given by the ratio
\begin{equation}
\label{eq:delta}
\delta_2(z) = \frac{r_\mathrm{s}}{[V_2(z)]^{1/3}}\,,
\end{equation}
where the comoving volume $V_2(z)$ is defined as
\begin{equation}
\label{eq:comvol}
V_2(z) = \frac{cz}{H_2(z)}\left[(1+z) D_2(z)\right]^2\,.
\end{equation}
From Eqs.~\eqref{eq:delta}--\eqref{eq:comvol}, it is clear that such measurements enable constraints on $\omega_b$ and $\omega_m$ from the definition of $r_\mathrm{s}$, on $h_0$ from $H_2(z)$, and on $\Omega_k$ from $D_2(z)$.

The second kind of BAO data (see Tab.~\ref{tab:BAO}) is a sample of $N_\Delta=4$ points described by the interpolated ratio
\begin{equation}
\label{eq:Delta}
\Delta_2(z) = \frac{D_2(z)}{r_\mathrm{s}}\,,
\end{equation}
that constrains $\omega_b$ and $\omega_m$ from the definition of $r_\mathrm{s}$, and $\Omega_k$ from $D_2(z)$.

The third set of $N_\Theta=2$ BAO measurements of Tab.~\ref{tab:BAO} involves the interpolated ratio
\begin{equation}
\label{eq:Theta}
\Theta_2(z) = \frac{c}{r_\mathrm{s} H_2(z)}\,,
\end{equation}
that, besides $\omega_b$ and $\omega_m$ got from $r_\mathrm{s}$, leads to the constraint on $h_0$ from $H_2(z)$.

Finally, to break the degeneracy between $\omega_b$ and $\omega_m$, we employ the $N_A=3$ correlated BAO measurements \cite{2012MNRAS.425..405B} described by the interpolated acoustic parameter
\begin{equation}
\label{eq:A}
A_2(z) \equiv g_\alpha\frac{\sqrt{\omega_m}}{cz}[V_2(z)]^{1/3}\,,
\end{equation}
that enables constraints on $\omega_m$, $h_0$, and $\Omega_k$.

With the usual assumption of Gaussian distributed errors $\sigma_{X_j}$, the log-likelihood functions of the uncorrelated BAO data sets ($X=\delta$, $\Delta$, $\Theta$) are given by
\begin{equation}
\label{loglikeBAOu}
    \ln \mathcal{L}_{\rm X} = -\frac{1}{2} \sum_{j=1}^{N_{\rm X}}\left\{\left[\dfrac{X_j-X_2(z_j)}{\sigma_{X_j}}\right]^2 + \ln(2\pi\sigma^2_{X_j})\right\}.
\end{equation}

Conversely, the log-likelihood function for correlated BAO data with covariance matrix $\mathbf{C}_{\rm B}$ \cite{2012MNRAS.425..405B} writes as
\begin{equation}
\label{loglikeBAOc}
\ln \mathcal{L}_{\rm A} = -\frac{1}{2} \left[\Delta{\bf A_2}^{\rm T} \mathbf{C}_{\rm B}^{-1}
\Delta{\bf A_2} + \ln \left(2 \pi |\det\mathbf{C}_{\rm B}| \right)\right]\,,
\end{equation}
where $\Delta{\bf A_2}\equiv A_j-A_2(z_j)$.

Finally, the total BAO log-likelihood can be written as
\begin{equation}
\label{loglikeBAO}
    \ln \mathcal{L}_{\rm B} = \sum_X\ln \mathcal{L}_{\rm X} + \ln \mathcal{L}_{\rm A}\,.
\end{equation}

\section{Numerical results}\label{sec3}

To set the bounds over $\omega_b$, $\omega_m$, $h_0$, and $\Omega_k$, we performed an MCMC analysis, based on the Metropolis-Hastings algorithm \cite{1953JChPh..21.1087M, 1970Bimka..57...97H}, through a modified version of the \texttt{Wolfram Mathematica} free available code presented in Ref.~\cite{2019PhRvD..99d3516A}.
We search for the best-fit results that maximize the total log-likelihood function
\begin{equation}
 \ln{\mathcal{L}} = \ln{\mathcal{L}_{\rm O}} + \ln{\mathcal{L}_{\rm S}} + \ln{\mathcal{L}_{\rm B}}\,,
\end{equation}
with the following priors on the parameters
\begin{equation}
\nonumber
\begin{array}{rclcrclr}
\alpha_0\equiv h_0 & \in & \left[0,1\right] &\qquad & \qquad \Omega_k & \in &\left[-2,2\right] &,\\
\alpha_1 & \in & \left[0,2\right] & \qquad & \qquad
\omega_b & \in & \left[0,1\right] &,\\
\alpha_2 & \in & \left[0,3\right] &\qquad &\qquad\,\,\, \omega_m & \in & \left[0,1\right] &.
\end{array}
\end{equation}

\begin{figure*}
\centering
\includegraphics[width=0.82\hsize,clip]{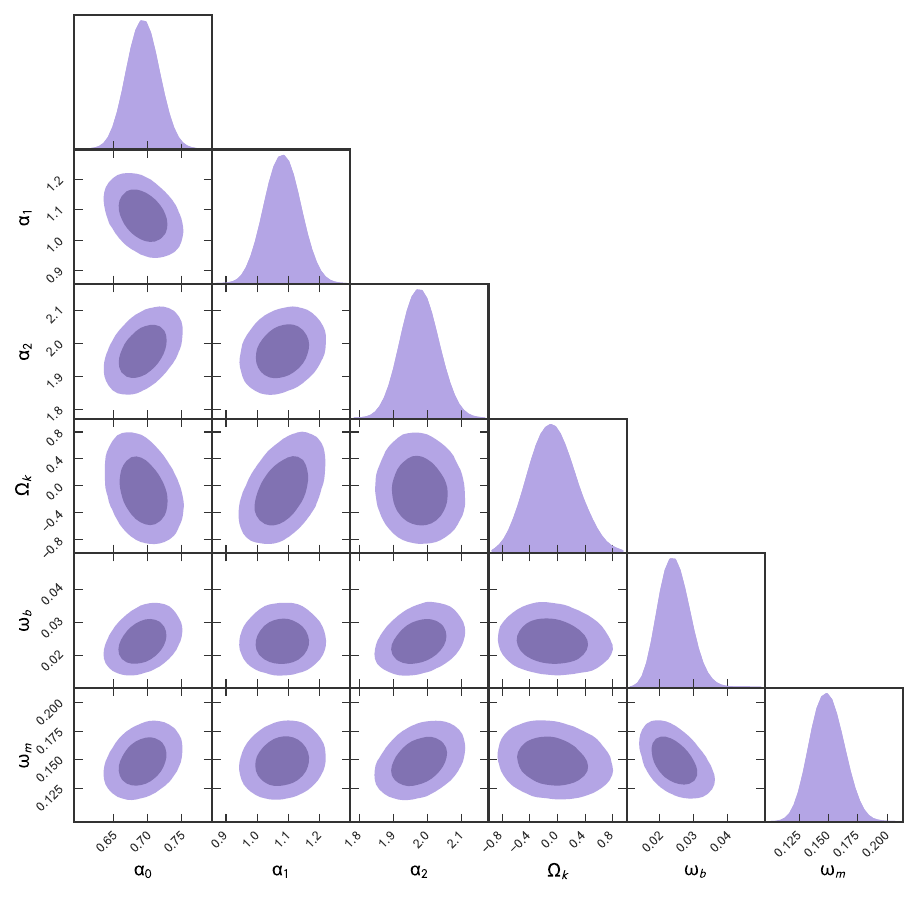}
\caption{MCMC contour plots of the B\'ezier interpolation. Darker (lighter) areas exhibit $1$--$\sigma$ ($2$--$\sigma$) confidence regions.}
\label{fig:cont}
\end{figure*}

\begin{figure}
\centering
\includegraphics[width=\hsize,clip]{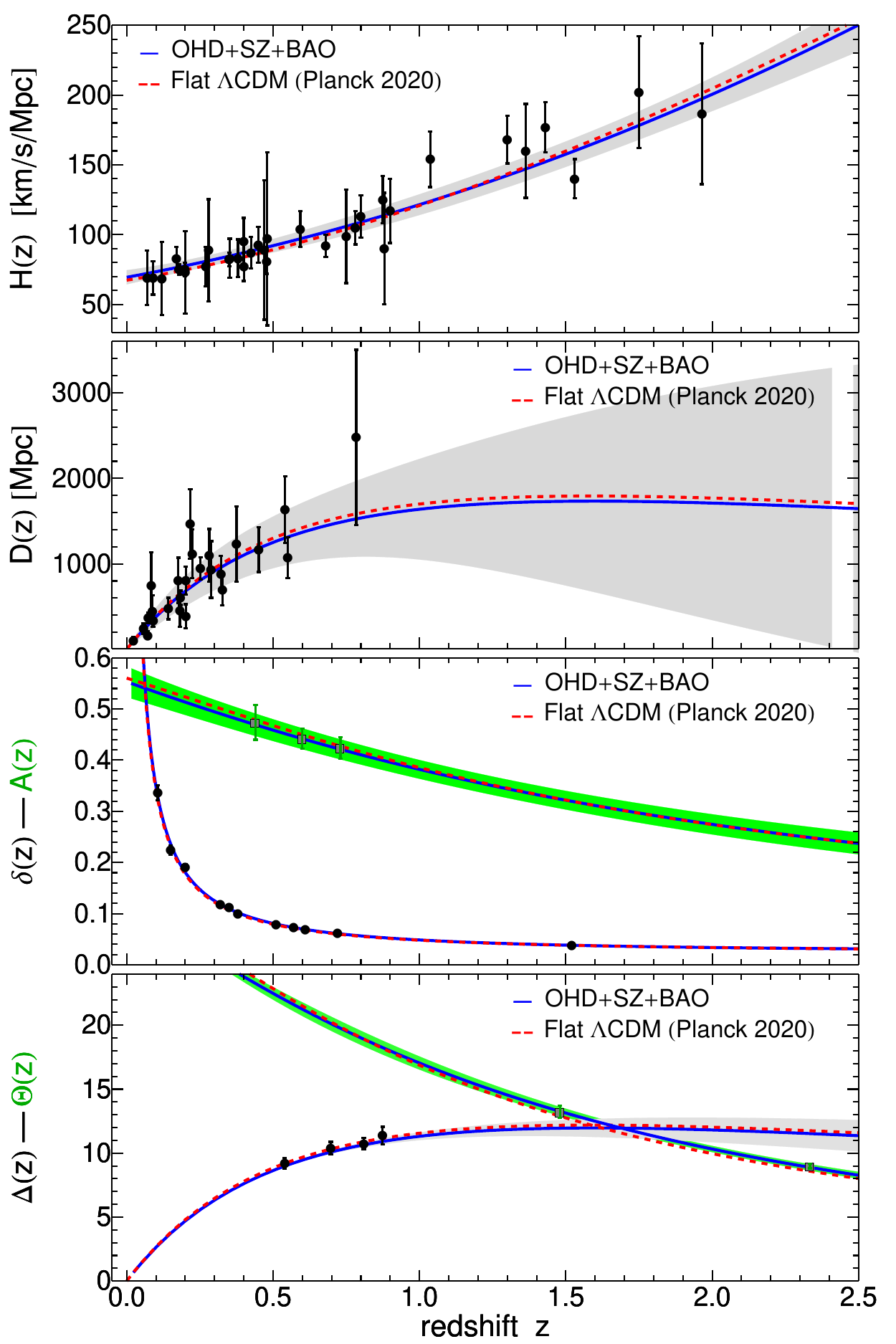}
\caption{Plots of the best-fitting B\'ezier curves (blue lines) and $1$--$\sigma$ confidence bands for $H_2(z)$, $D_2(z)$, $\delta_2(z)$, and $\Delta_2(z)$ (gray bands), and $A_2(z)$ and $\Theta_2(z)$ (green bands), compared to the $\Lambda$CDM paradigm \citep{Planck2018} (dashed red curves).}
\label{fig:Bez}
\end{figure}

The best-fit parameters of our MCMC analysis are listed in Tab.~\ref{tab:bestfit} and portrayed in the $1$--$\sigma$ and $2$--$\sigma$ contour plots of Fig.~\ref{fig:cont}, obtained by using a Python free available code \cite{2016JOSS....1...46B}.
For comparison, Tab.~\ref{tab:bestfit} also lists the flat and non-flat $\Lambda$CDM best-fit parameters got from the CMB measurements \cite{Planck2018}.
The best-fitting B\'ezier curves that interpolate OHD, SZ, and BAO catalogs are portrayed in Fig.~\ref{fig:Bez}, where the flat $\Lambda$CDM paradigm \citep{Planck2018} is also shown for comparison.

\begin{table*}
\centering
\setlength{\tabcolsep}{0.5em}
\renewcommand{\arraystretch}{1.3}
\begin{tabular}{lccccc}
\hline\hline
                                &
$\alpha_0\equiv h_0$            &
$\alpha_{1,2}$                  &
$\Omega_k$                      &
$\omega_b$                      &
$\omega_m$\\
\hline
This work                       &
$0.694_{-0.035\,(-0.055)}^{+0.032\,(+0.054)}$ &
$1.073_{-0.077\,(-0.123)}^{+0.088\,(+0.140)}$ &
$-0.08_{-0.52\,(-0.77)}^{+0.52\,(+0.84)}$     &
$0.0236_{-0.0058\,(-0.0089)}^{+0.0069\,(+0.0118)}$ &
$0.147_{-0.020\,(-0.031)}^{+0.021\,(+0.034)}$\\
  &  & $1.973_{-0.076\,(-0.122)}^{+0.079\,(+0.132)}$ &  &\\
$\Lambda$CDM                    &
$0.6736^{+0.0054}_{-0.0054}$    &
--                              &
--                              &
$ 0.02237^{+0.00015}_{-0.00015}$&
$0.1430^{+0.0011}_{-0.0011}$ \\
$\Lambda$CDM (non-flat)  &
$0.636^{+0.021}_{-0.023}$       &
--                              &
$-0.011^{+0.013}_{-0.012}$      &
$0.02249^{+0.00016}_{-0.00016}$ &
$0.14099^{+0.0015}_{-0.0015}$ \\
\hline
\end{tabular}
\caption{Best-fit parameters with $1$--$\sigma$ ($2$--$\sigma$) error bars obtained from the MCMC analysis based on the B\'ezier interpolation of this work, compared to the flat and non-flat $\Lambda$CDM best-fit parameters \cite{Planck2018}. In the non-flat $\Lambda$CDM case: a) $\Omega_k$ is quoted at $95\%$ limits, whereas the other parameters at $68\%$ limits \cite{Planck2018}; b) the value of $\omega_m$ is calculated by summing up $\omega_b$ with the density of cold dark matter, \textit{i.e.} $\omega_c = 0.1185^{+0.0015}_{-0.0015}$, and the attached error as the root sum of squares of the errors on $\omega_b$ and $\omega_c$ \cite{Planck2018}.}
\label{tab:bestfit}
\end{table*}

The above results evidence that our interpolation technique of intermediate-redshift data sets is able to set precise model-independent constraints over the key cosmological parameters $\omega_b$, $\omega_m$, $h_0$, and $\Omega_k$ listed in Tab.~\ref{tab:bestfit}. It is immediately clear that these cosmic bounds are in agreement within $1$--$\sigma$ ($2$--$\sigma$) with those got from the flat (non-flat) concordance model, though with larger attached errors.
In particular, the Hubble constant tension remains still unsolved, since our value of $h_0$ listed in Tab.~\ref{tab:bestfit} is still consistent at $1$--$\sigma$ confidence level with the value got from the CMB data ($h_0 = 0.6736\pm 0.0054$) \cite{Planck2018} and from SNe Ia ($h_0=0.7304\pm0.0104$) \cite{2022ApJ...934L...7R}, both obtained in the flat scenario.
In the non-flat case, our $h_0$ is consistent with Planck only at $2$--$\sigma$ confidence level \cite{Planck2018}.

The spatial curvature found in this work (see Tab.~\ref{tab:bestfit})
is compatible at $2$--$\sigma$ level with the flat geometry purported by {\it Planck} \cite{Planck2018}. However, the large attached error does not to exclude {\it a priori} other geometries. Such a loose constraint is due to scattered and limited redshift span (with respect to the other catalogs here employed) of the SZ data (see second panel in Fig.~\ref{fig:Bez}).

\section{Final outlooks and perspectives}
\label{sec4}

In this paper, we proposed a novel model-independent technique to extract bounds on the key cosmological parameters, such as the normalized Hubble constant $h_0$, the curvature parameter $\Omega_k$ and the matter densities $\omega_b$, for baryons, and $\omega_m$, for all the matter components.

To do so, we resorted a calibration technique based on the well-established \emph{B\'ezier parametric curve} and apply it to interpolate the OHD catalog with a second order polynomial curve \cite{2019MNRAS.486L..46A,LM2020,2021MNRAS.501.3515M,2023MNRAS.518.2247L,2023MNRAS.523.4938M}, without assuming any \emph{a priori} cosmological models.
Though no assumptions are made, by construction, the interpolating function, $H_2(z)$, carries out the valuable constraint on $\alpha_0\equiv h_0$.

Afterwards, we interpolated other intermediate redshift catalogs, based on the SZ effect measurements and BAO data sets.
Recently, SZ data have been used in conjunction with SNe Ia to infer cosmic bounds on $H_0$ through the cosmic distance duality
relation \cite{2023arXiv231018711C}.
Here, we intentionally excluded SNe Ia \cite{2018ApJ...859..101S} and also CMB data \cite{Planck2018}, mainly because of the existing tension on the values of $h_0$ got from these two probes.

Conversely, we used the Hubble rate $H_2(z)$ to get an interpolated angular diameter distance $D_2(z)$ which can be compared with SZ measurements.
Since the interpolations $H_2(z)$ and $D_2(z)$ do not bear \emph{a priori} assumptions on the spatial curvature, as long as the data sets do not carry specific priors on it, SZ data provided model-independent bounds on $\Omega_k$.

Contrary to the procedure we worked out in Ref.~\cite{2023MNRAS.518.2247L}, where OHD and BAO catalogs were both interpolated by means of two different B\'ezier parametric curves, here we used four different data sets of BAO measurements and compared them with the interpolated function obtained by the combinations of the above determined $H_2(z)$ and $D_2(z)$, aiming to provide bounds on $\omega_b$ and $\omega_m$ and to reinforce the constraints on $h_0$ and $\Omega_k$, previously obtained from OHD and SZ data sets, respectively.

Specifically, the constraints on $\omega_b$ and $\omega_m$ are derived from the comoving sound horizon, $r_{\rm s}$, on which BAO data depend and, in general, $\omega_b$ is fixed to the value got from the CMB. Since we do not employ   CMB data, here the degeneracy between $\omega_b$ and $\omega_m$, exhibited by Eq.~\eqref{eq:neutrino}, is bypassed  utilizing the interpolated acoustic parameter, $A_2(z)$, for correlated BAO measurements \cite{2012MNRAS.425..405B}.

The results provided by the MCMC analysis, based on the Metropolis algorithm and shown in Tab.~\ref{tab:bestfit} and Figs.~\ref{fig:cont} and \ref{fig:Bez}, evidence that our model-independent technique provides precise constraints, though with larger attached errors, over the key cosmological parameters. These constraints are in agreement within $2$--$\sigma$ level with those got from the non-flat extension of the concordance model, and are in a better agreement, within $1$--$\sigma$, also with $h_0$ got from the flat scenario of the $\Lambda$CDM model.

The $h_0$ tension is still not fully-addressed, even though OHD+SZ+BAO provide narrow constraints, see Tab.~\ref{tab:bestfit}.
Indeed, at $1$--$\sigma$ confidence level, our $h_0$ appears more consistent with Planck estimates in the flat scenario \cite{Planck2018} and barely consistent with SNe Ia, i.e., $h_0=0.7304\pm0.0104$ \cite{2022ApJ...934L...7R}.

Accordingly, it is worth mentioning that a recent estimate got from SNe Ia based on surface brightness fluctuations measurements, i.e., $h_0=0.7050\pm0.0237$ \cite{2021A&A...647A..72K}, not only agrees with our findings, but seems also to indicate that the Hubble constant may be in between the extreme values.

Last but not least, when compared to the non-flat case of the $\Lambda$CDM model, our $h_0$ is consistent with Planck only at $2$--$\sigma$ confidence level \cite{Planck2018}, as prompted in Tab.~\ref{tab:bestfit}.

Our overall outputs on the spatial curvature, see Tab.~\ref{tab:bestfit}, is thus compatible at $2$--$\sigma$ level with the flat geometry purported by the $\Lambda$CDM model and with its non-flat extension \cite{Planck2018}. In this respect, though no tension with the concordance model manifestly arose, our bounds on $\Omega_k$ cannot exclude {\it a priori} non-flat geometries, only likely less probable than the flat case.
This loose constraint is mainly due to scattered and limited redshift span of the SZ data, as one can notice from the second panel of Fig.~\ref{fig:Bez}.

We conclude that our model-independent method provides accurate constraints confirming the $\Lambda$CDM background. Hence, neither minimal extensions of the standard cosmological model \citep{Izzo2012c,2021ApJ...908..181M,Luongo:2021nqh} nor additional terms in the Hilbert-Einstein action \cite{Capozziello:2019cav} seem to be required.

To further refine our constraints, it would be crucial for the future to enlarge and improve the quality of the catalogs involved in this analysis, in particular the SZ data that significantly impact on the $\Omega_k$ estimate.

Finally, we remark that our technique can be used to calibrate GRB \cite{2021Galax...9...77L} and quasar \cite{2019NatAs...3..272R} correlations in a model-independent way, as proposed in Ref.~\cite{2023MNRAS.518.2247L}, strengthening the constraints, and pushing the farther our analysis up to $z\sim9$. Thus, new intermediate data catalogs to explore with the same treatment here-described will be object of future efforts.

\section*{Acknowledgements}
The work of OL and MM is partially supported by the Ministry of Education and Science of the Republic of Kazakhstan, Grant IRN AP08052311.

%

\end{document}